\documentclass[10pt,letterpaper]{article}
\usepackage{opex3}
\usepackage{graphicx} 

\begin{document}



\title{
Carrier-carrier scattering and negative dynamic conductivity in pumped graphene \textit{}}

\author{Dmitry Svintsov,$^{1,2,3*}$ Victor Ryzhii,$^{1,4}$ Akira Satou,$^{1}$ Taiichi Otsuji,${^1}$ and Vladimir Vyurkov$^{2}$}
\address{$^1$ Research Institute for Electrical Communication, Tohoku University, Sendai 980-8577, Japan\\ 
$^2$ Institute of Physics and Technology, Russian Academy of Sciences, Moscow 117218, Russia\\
$^3$ Department of General 
Physics, Moscow Institute of Physics and Technology,
Dolgoprudny 141700, Russia\\
$^4$ Center for Photonics and Infrared Engineering, Bauman Moscow State Technical University,
Moscow 105005, Russia
}

\email{*svintcov.da@mipt.ru} 



\begin{abstract} 
We theoretically examine the effect of carrier-carrier scattering processes (electron-hole and electron-electron) on the intraband radiation absorption and their contribution to the net dynamic conductivity in optically or electrically pumped graphene. We demonstrate that the radiation absorption assisted by the carrier-carrier scattering can be stronger than the Drude absorption due to the carrier scattering on disorder.
Since the intraband absorption of radiation effectively competes with its interband amplification,
this can substantially affect the conditions of the negative dynamic conductivity in
the pumped graphene and, hence, the interband terahertz and infrared lasing.
We find the threshold values of the frequency and quasi-Fermi energy of nonequilibrium carriers corresponding to the onset of negative dynamic conductivity. The obtained results show that the effect of carrier-carrier scattering shifts the threshold frequency of the radiation amplification in pumped graphene to higher values.
In particular, the negative dynamic conductivity is attainable at the frequencies above $6$ THz in graphene on SiO$_2$ substrates at room temperature. The threshold frequency can be decreased to markedly lower values in graphene structures with high-$\kappa$ substrates due to screening of the carrier-carrier scattering, particularly at lower temperatures. 
\end{abstract}

\ocis{(160.4760) Optical properties; (160.6000) Semiconductor materials; (140.3070) Infrared and far-infrared lasers; (250.0250) Optoelectronics.} 


\newpage
\section{Introduction}

Graphene, a two-dimensional carbon crystal, possesses no energy band gap and, hence, is promising for detection and generation of far-infrared and terahertz 
(THz) radiation~\cite{Graphene-photonics-review-1,Graphene-photonics-review-2,Pumped-graphene-materials}. Several concepts of graphene-based THz lasers with optical pumping~\cite{Optical-pumping-laser} or electrical (injection) pumping~\cite{Electrical-pumping-laser1,Electrical-pumping-laser2} have been proposed and analyzed. Recently, the possibility of THz-wave amplification by optically pumped graphene was shown experimentally~\cite{THz-amplification-experim}. 

Generally, one of the main reasons for low efficiency of THz semiconductor laser is the intraband (Drude) radiation absorption~\cite{THz-gain-loss-QCL}, which grows rapidly with decreasing the radiation frequency. This absorption process aggressively competes with the radiation amplification due to the stimulated interband electron transitions under the  conditions of population inversion. This problem is crucial for graphene-based lasers as well.

\begin{figure}[t]
\center{\includegraphics[width=0.95\linewidth]{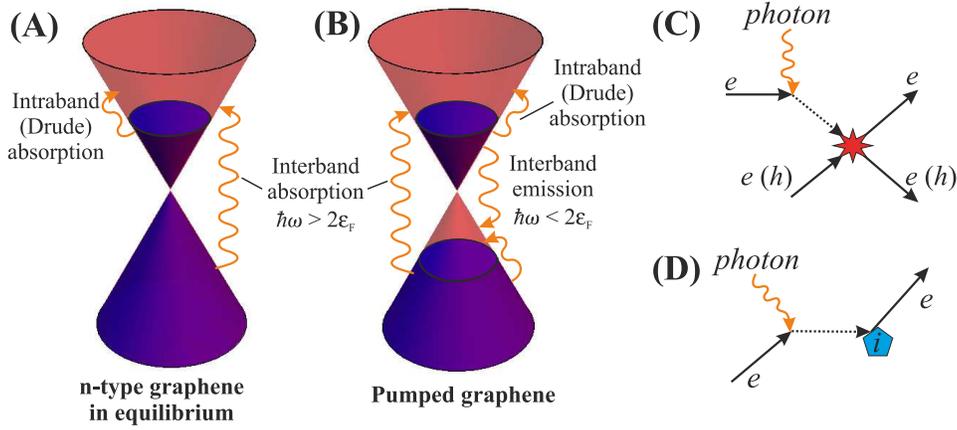}}
\caption{Schematic views of band diagrams of (A) n-type graphene in equilibrium and (B) pumped graphene. Wavy arrows indicate photon absorption and emission processes. Diagram of photon absorption by an electron ({e}) associated with (C) electron-electron (electron-hole) and (D) impurity ({i}) scattering.}
\label{F1}
\end{figure}

The optical conductivity of clean undoped graphene is equal to the universal value $\sigma_q = e^2/4\hbar$, where $e$ is the elementary charge and $\hbar$ is the Planck's constant. This value corresponds to the interband radiation absorption coefficient $ \pi \alpha =2.3$ \%~\cite{Optical-conductivity-measurement}, where $\alpha = e^2/\hbar c$ is the fine-structure constant ($c$ is the speed of light in vacuum). The optical conductivity of pumped graphene can be negative due to prevailing stimulated electron transitions from the conduction to the valence band associated with the interband population inversion~\cite{Negative-dynamic-conductivity}. The interband dynamic conductivity appears to be negative at frequencies $\omega < 2\epsilon_F/\hbar$, where $\epsilon_F$ is the quasi-Fermi energy of pumped carriers [see Fig.~\ref{F1} (A) and (B)]. However, the coefficient of radiation amplification in pumped graphene is still limited by $2.3$ \%. Hence, for the operation of graphene-based lasers the Drude absorption coefficient should lie below $2.3$ \%. In the previous models of negative dynamic conductivity in pumped graphene,
it was usually assumed that the Drude absorption originates from electron and hole scattering on impurities, lattice defects, and phonons~[Fig.~\ref{F1} (D)], which can, in principle, be removed almost completely in high-quality graphene samples and at low temperatures (see, for example, Refs. ~\cite{Pumped-graphene-materials, Threhold-of-population-inversion}). However, there is an unavoidable mechanism leading to the intraband absorption, namely, the radiation absorption assisted by the carrier-carrier scattering~[Fig.~\ref{F1} (C)].

The carrier-carrier scattering (for brevity denoted as c-c scattering) was shown to be a key factor in the relaxation kinetics of photoexcited electrons and holes in graphene~\cite{THz-amplification-experim,Collinear-Scattering,Relaxation-kinetics}. It can also be responsible for weakly temperature-dependent minimal dc conductivity of graphene~\cite{Kashuba,Quantum-critical,Our-hydrodynamics,Interactions-transport,Electron-interactions}. The strength of c-c scattering in graphene is governed by the relatively large coupling constant, $\alpha_c = e^2/(\kappa_0 \hbar v_F)\sim 1$, where $v_F=10^6$ m/s is the velocity of massless electrons and holes (Fermi velocity) and $\kappa_0$ is the background dielectric constant. 

Among a variety of phenomena originating from the carrier-carrier scattering in graphene-based structures, there are modifications of quasiparticle spectra~\cite{Electron-interactions}, relaxation~\cite{Relaxation-kinetics}, thermalization, and recombination~\cite{Collinear-Scattering}. In this paper,  we focus   on the scattering-assisted intraband radiation absorption. The latter appears to be a quite strong absorption mechanism, especially in the pumped graphene with population inversion. We derive the expression for the real part of the intraband dynamic conductivity $\mathrm{Re}\sigma_{cc}$ (which is proportional to the intraband contribution to the radiation absorption) arising due to carrier-carrier collisions in clean pumped graphene. The probability of corresponding intraband radiation absorption process is evaluated using the second-order perturbation theory and the Fermi golden rule. Comparing $\mathrm{Re}\sigma_{cc}$ with the interband conductivity $\mathrm{Re}\sigma_{inter}$, we find the threshold values of the quasi-Fermi energies of pumped carriers, required to attain the net negative dynamic conductivity (negative absorption coefficient) at given frequency, and calculate the pertinent threshold frequencies. We also show that screening of the Coulomb potential by the carriers in graphene plays the significant role in the c-c scattering. Because of screening, the intraband conductivity due to the c-c collisions grows slowly (almost linearly) with increasing the quasi-Fermi energy of pumped carriers. We find that in the pumped graphene, the main contribution to the intraband conductivity arises from electron-hole (e-h) scattering, while the electron-electron (e-e) and hole-hole (h-h) collisions yield less than one tenth of its total value. The rate of c-c scattering and the corresponding radiation absorption can be reduced in graphene clad between materials with high dielectric constants.

The paper is organized as follows. In Sec.~2, we derive the general equations for the radiation power absorbed due to c-c scattering-assisted intraband transitions, and find the pertinent contribution to the dynamic conductivity. In Sec.~3, we obtain the dependences of the intraband, interband,
and net dynamic conductivity on the radiation frequency, the quasi-Fermi energy, the background dielectric constant; and find the frequency threshold of the radiation amplification.
Section~4 deals with discussion of the obtained results and their validity.
In Sec.~5, we draw the main conclusions.
Some mathematical details are singled out in Appendix.

\section{Intraband dynamic conductivity. General equations} 

There are two mechanisms through which the c-c scattering affects the optical conductivity, and, hence, absorption of radiation. First, the finite lifetime of carriers leads simply to the smearing of the interband absorption/amplification edge~\cite{Interplay-of-inter-and-intra}. Second, a free electron can absorb a photon and transfer the excess energy and momentum to the other carrier [Fig.~\ref{F1} (C)]. The latter mechanism is quite similar to the conventional Drude absorption, where the excess momentum is transferred to an impurity or phonon [Fig.~\ref{F1} (D)]. However, in the case of graphene, the radiation absorption due to c-c scattering exhibits a number of unusual features.

In semiconductors with parabolic bands, e-e scattering cannot directly affect the conductivity and absorption of radiation. The reason is that the total current carried by two electrons is not changed in the scattering process due to momentum conservation. In graphene, however, electron velocity and momentum are not directly proportional to each other (sometimes referred to as ''momentum-velocity decoupling''~\cite{Interactions-transport}). Hence, the momentum conservation does not imply the current conservation, and e-e scattering can contribute to optical absorption. Such effects were studied in III-V semiconductors in the light of $p^4$-corrections to the parabolic bands~\cite{EE-absorption-in-QW}. In graphene, the carrier dispersion law is not parabolic ''from the very beginning''. Hence, those effects can play a more considerable role.

Electron-hole (e-h) scattering processes do not conserve the total current and can contribute to radiation absorption independent of the carrier spectrum. Apart from graphene with Fermi energy close to Dirac point, e-h scattering plays a minor role in conductivity due to vanishing number of holes. This is not the case of the pumped graphene, where e-h scattering is intensified due to the great number of carriers in both bands.

We consider a two-step quantum-mechanical process: the quasiparticle (electron or hole) absorbs light quantum passing to the virtual state, then it collides with other quasiparticle and transfers the excess energy (and momentum) to it. If the frequency of electromagnetic wave $\omega$ substantially exceeds the electron (or hole) collision frequency $\nu$, one can use the Fermi golden rule to calculate the absorption probability in unit time. We will first consider the radiation absorption due to e-e collisions, and then generalize the result to account for e-h processes. The initial momenta of incoming quasiparticles are denoted by ${\bf p}_1$ and ${\bf p}_2$, while the final momenta are ${\bf p}_3$ and ${\bf p}_4$.

To work out the matrix elements appearing in the Fermi golden rule we apply the second order perturbation theory considering the two perturbations. The first one, $\hat V_F (t)$, is the interaction of electron with external electromagnetic field

\begin{equation}
\hat V_F (t) = \frac{e}{2c}( {\bf A}_0, \hat {\bf v}_1 + \hat {\bf v}_2)(e^{i \omega t} + e^{-i \omega t}),
\end{equation}
where $\hat{\bf v}_1$ and $\hat {\bf v}_2$ are the velocity operators of the two electrons, ${\bf A}_0$ is the amplitude of field vector potential, and $\omega$ is the frequency of the electromagnetic wave. The vector ${\bf A}_0$ and the corresponding electric field ${\bf E}_0 = i\omega {\bf A}_0/c$ lie in the plane of graphene layer.

The second perturbation, $\hat V_C$, is the Coulomb interaction between carriers. It has nonzero matrix elements connecting the two-particle states $| {\bf p}_1 , {\bf p}_2 \rangle$ and $| {\bf p}_3, {\bf p}_4 \rangle = | {\bf p}_1 + {\bf q}, {\bf p}_2 - {\bf q}\rangle$, where ${\bf q}$ is the transferred momentum. We denote these matrix elements by $V_C({\bf q})\equiv \langle {\bf p}_3, {\bf p}_4 | \hat{V}_C | {\bf p}_1, {\bf p}_2 \rangle $, 
\begin{equation}
V_C({\bf q}) = \frac{2\pi e^2 \hbar }{q \kappa_0 \kappa (q)}\left\langle u^{(e)}_{{\bf p}_1} u^{(e)}_{{\bf p}_3} \right\rangle \left\langle u^{(e)}_{{\bf p}_2} u^{(e)}_{{\bf p}_4} \right\rangle,
\end{equation}
where $\langle {u^{(e)}_{{\bf p}_i}}{u^{(e)}_{{\bf p}_j}} \rangle =\cos (\theta_{ij}/2)$ are overlap factors of the electron envelope functions in graphene, $\theta_{ij}$ is the angle between momenta ${\bf p}_i$ and ${\bf p}_j$, $\kappa_0$ is the background dielectric constant, and $\kappa ( q )$ is the dielectric function of graphene itself. For our purposes, it can be taken in the static limit

\begin{equation}
\kappa(q) = 1 + q_{TF}/q,
\end{equation}
$q_{TF}$ is the Thomas-Fermi (screening) momentum~\cite{Hwang-Das-Sarma}.

A well-known relation for the second-order matrix element connecting initial $|i\rangle = | {\bf p}_1, {\bf p}_2 \rangle$ and final $|f\rangle=| {\bf p}_3, {\bf p}_4 \rangle$ states reads~\cite{LL-Quantum-mechanics}

\begin{equation}
\langle f |\hat V | i \rangle =\sum\limits_{m}{\frac{ \langle f | \hat V_{F\omega}| m \rangle \langle m | \hat V_C | i \rangle } {\left( \varepsilon_{{\bf p }_1} + \varepsilon_{{\bf p }_2} \right)-\left( \varepsilon_{{\bf p }_{1m}}+\varepsilon_{{\bf p }_{2m}} \right)}} + 
\frac{ \langle f | \hat V_{C} | m \rangle \langle m | \hat V_{F\omega}| i \rangle } {\left( \varepsilon_{{\bf p }_1} +\varepsilon_{{\bf p }_2}+\hbar \omega \right)-\left( \varepsilon_{{\bf p }_{1m}}+\varepsilon_{{\bf p }_{2m}} \right)},
\end{equation}
where the index $m$ counts the intermediate states, $\varepsilon_{\bf p}=p v_F$ is the electron dispersion law in graphene, and $\hat V_{F\omega} = (e/2c)\left( {\bf A}_0, {{\bf{\hat v}}_1}+{{\bf{\hat v}}_2} \right)$ is the Fourier-component of electron-field interaction. The summation over intermediate momenta ${\bf p }_{m}$ is easily performed as photon momentum is negligible compared to electron momentum. Using also the energy conservation law

\begin{equation}
\varepsilon_{{\bf p }_1}+\varepsilon_{{\bf p }_2}+\hbar \omega = \varepsilon_{{\bf p }_3}+\varepsilon_{{\bf p }_4},
\end{equation} 
we obtain a very simple relation for matrix element

\begin{equation}
\label{Vfi}
\langle f | V | i \rangle = \frac{e}{2c} \frac{V_C ({\bf q})}{\hbar \omega } \left( {\bf A}_0, {\bf v}_{{\bf p }_1}+{\bf v}_{{\bf p }_2}-{\bf v}_{{\bf p}_3}-{\bf v}_{{\bf p }_4} \right) \cos (\theta_{13}/2 ) \cos (\theta_{24}/2).
\end{equation}

Equation (\ref{Vfi}) clearly demonstrates that for parabolic bands no absorption due to e-e scattering can occur as ${\bf v}_{{\bf p }_1}+{\bf v}_{{\bf p }_2}-{\bf v}_{{\bf p }_3}-{\bf v}_{{\bf p }_4}$ turns to zero as a result of momentum conservation.

The power, $P$, absorbed by an electron system is calculated using the Fermi golden rule, taking into account the occupation numbers of the initial and final states:
$$
P =\frac{\hbar \omega} {2} \frac{2\pi g^2}{\hbar } \sum\limits_{{\bf p }_1,{\bf p }_2, {\bf q}}{|\langle f | V | i \rangle|^2}
\delta \left( \varepsilon_{{\bf p }_1} + \varepsilon_{{\bf p }_2} +\hbar \omega - \varepsilon_{{\bf p}_3} - \varepsilon_{{\bf p}_4} \right)
$$
\begin{equation}
\label{dWdt}
\times f_e\left({\bf p}_1 \right) f_e\left( {\bf p}_2 \right)\left[ 1-f_e\left( {\bf p}_3 \right) \right] \left[ 1-f_e\left( {\bf p}_4 \right) \right][ 1-\exp(-\hbar \omega/k_BT)].
\end{equation}
Here $g=4$ stands for the spin-valley degeneracy factor in graphene. The pre-factor $1/2$ cuts off the equivalent scattering processes; without it the indistinguishable collisions ${\bf p}_1 + {\bf p}_2$ and ${\bf p}_2 + {\bf p}_1$ are treated separately, which would be incorrect. The function $f_e({\bf p})$ in Eq.~(\ref{dWdt}) is the electron distribution function, which is assumed to be the quasi-equilibrium Fermi function:

\begin{equation}
f_e({\bf p}) = \left[1+ \exp\left(\frac{\varepsilon_{\bf p} - \mu_e}{k_B T}\right)\right]^{-1}.
\end{equation}
To treat the radiation absorption in pumped graphene, we introduce the different quasi-Fermi energies of electrons and holes, $\mu_e$ and $\mu_h$. In symmetrically pumped systems $\mu_e = -\mu_h = \epsilon_F > 0$. In what follows, if otherwise not stated, we will consider symmetrically pumped graphene. In such system, the occupation numbers of electrons and holes with energies $\varepsilon_{\bf p} > 0$ are equal, hence, the subscripts $e$ and $h$ of the distribution functions can be omitted. In deriving Eq.~(\ref{dWdt}) we have also neglected the exchange-type scattering, which occurs only between electrons with same spin and from the same valley. This assumption does not significantly affect the final numerical values. 

Finally, to obtain the real part of the intraband dynamic conductivity due to e-e collisions $\mathrm{Re}\sigma_{ee}$, we express the vector-potential in Eq.~(\ref{Vfi}) through electric field and equate (\ref{dWdt}) with $\mathrm{Re}\sigma_{ee} E_0^2/2$. As a result, this conductivity is expressed via the universal optical conductivity of clean graphene $\sigma_q $, the coupling constant $\alpha_c = e^2/(\kappa_0 \hbar v_F)$, and the dimensionless 'collision integral' $I_{ee,\omega}$:

\begin{equation}
\label{Kee}
\mathrm{Re}\sigma_{ee}=\sigma_q \frac{\alpha_c^2}{\pi^3} \left( \frac{k_B T}{\hbar \omega } \right)^3 \left[ 1-\exp\left(-\frac{\hbar \omega }{k_B T}\right) \right] I_{ee,\omega},
\end{equation}
$$
I_{ee,\omega} = \int{\frac{ d{\bf Q} d {\bf k}_1 d {\bf k}_2} {Q^2 \kappa^2(Q)} (\Delta {\bf n}_{ee})^2 \cos^2\left(\theta_{1\pm}/2\right) \cos^2\left(\theta_{2\pm}/2 \right) } 
$$
\begin{equation}
\label{Iee}
\times\delta[ k_{1+} + k_{2-} + \hbar \omega /(k_B T) - k_{1-} - k_{2+}] 
F ( k_{1+} ) F( k_{2-} ) \left[ 1- F( k_{1-}) \right] \left[ 1 - F( k_{2+}) \right].
\end{equation}
Here, for the simplicity of further analysis, we have introduced the dimensionless momenta ${\bf k}_{1\pm} =( {\bf p}_1 \pm {\bf q}/2) v_F /(k_B T) $, ${\bf k}_{2\pm} =( {\bf p}_2 \pm {\bf q}/2) v_F /(k_B T) $, $Q = {\bf q} v_F / (k_B T)$, as well as the dimensionless change in the electron current $\Delta {\bf n}_{ee} = ({\bf k}_{1+}/k_{1+} + {\bf k}_{2-}/k_{2-}) - ({\bf k}_{1-}/k_{1-} + {\bf k}_{2+}/k_{2+})$; 
$\theta_{1\pm}$ and $\theta_{2\pm}$ are the angles between ${\bf k}_{1+}$ and ${\bf k}_{1-}$
and between ${\bf k}_{2+}$ and ${\bf k}_{2-}$, respectively,
$F(x) = \{ 1+\exp[x - \epsilon_F/(k_B T)]\}^{-1}$ is the Fermi function of dimensionless argument.

At the frequencies exceeding the carrier collision frequency $\omega \gg \nu$, the contributions to the real part of dynamic conductivity from different scattering mechanisms are summed up. The intraband conductivity due to e-h collisions is given by expression similar to (\ref{Kee}-\ref{Iee}), with several differences. First, the change in current carried by electron and hole is given by $\Delta {\bf n}_{eh} = ({\bf k}_{1+}/k_{1+} - {\bf k}_{2-}/k_{2-}) - ({\bf k}_{1-}/k_{1-} - {\bf k}_{2+}/k_{2+})$ as the charge of hole is opposite to that of electron. Second, along with 'simple' electron-hole scattering, the annihilation-type interaction between electron and hole is also possible. In such process, electron and hole annihilate, emit a virtual photon, which produces again an electron-hole pair. The probability of such process is the same as of scattering process, but the overlap factors $\langle {u^{(e)}_{{\bf p}_i}}{u^{(e)}_{{\bf p}_j}}\rangle =\cos (\theta_{ij}/2)$ should be changed to $\langle {u^{(e)}_{{\bf p}_i}}{u^{(h)}_{{\bf p}_j}}\rangle =\sin (\theta_{ij}/2)$~\cite{Quantum-critical}. As a result, the electron-hole scattering contribution to the optical conductivity becomes

\begin{equation}
\label{Keh}
\mathrm{Re} {\sigma}_{eh}= \sigma_q \frac{2 \alpha_c^2}{\pi^3} \left( \frac{k_B T}{\hbar \omega } \right)^3 \left[ 1-\exp\left(-\frac{\hbar \omega }{k_B T}\right) \right] I_{eh,\omega},
\end{equation}
$$
I_{eh,\omega} = \int{\frac{ d{\bf Q} d {\bf k}_1 d {\bf k}_2} { Q^2 \kappa^2} (\Delta {\bf n}_{eh})^2
\left[ \cos^2\left(\theta_{1\pm}/2\right) \cos^2\left(\theta_{2\pm}/2 \right) + \sin^2\left(\theta_{1\pm}/2\right) \sin^2\left(\theta_{2\pm}/2 \right) \right]}\\
$$
\begin{equation}
\label{Ieh}
\times
\delta[ k_{1+} + k_{2-} + \hbar \omega /(k_B T) - k_{1-} - k_{2+}] F( k_{1+} ) F( k_{2-} ) \left[ 1 - F( k_{1-}) \right] \left[ 1 - F( k_{2+}) \right].
\end{equation}

Accordingly, the net intraband conductivity due to c-c collisions $\mathrm{Re}{\sigma}_{intra}$ in symmetrically pumped graphene is
\begin{equation}
\label{Ktot}
\mathrm{Re}{\sigma}_{cc} = \sigma_q \frac{2\alpha_c^2}{\pi^3} \left( \frac{k_B T}{\hbar \omega } \right)^3 \left[ 1-\exp\left(-\frac{\hbar \omega }{k_B T}\right) \right] \left( I_{ee,\omega} + I_{eh,\omega} \right).
\end{equation}
Here we have noted that h-h scattering contribution equals e-e contribution. The coefficient of radiation absorption is readily obtained from Eq.~(\ref{Ktot}) timing it by $4\pi/c$. The dimensionless 'collision integrals' $I_{ee,\omega}$ and $I_{eh,\omega}$ are given by Eqs.~(\ref{Iee}) and (\ref{Ieh}). They can be evaluated numerically passing to the elliptic coordinates~\cite{Collinear-Scattering}. In those coordinates, the delta-function can be analytically integrated, reducing the dimensionality of integral by unity. This procedure is described in Appendix.

\section{Analysis of intraband and net dynamic conductivity}

It is natural that the obtained expression~(\ref{Ktot}) for the real part of conductivity is proportional to the universal ac conductivity of clean undoped graphene $\sigma_q $, and the square of the coupling constant $\alpha_c$. The latter characterizes the strength of Coulomb interaction between carriers in graphene. The frequency behavior of $\mathrm{Re} \sigma_{cc}$ follows the well-known Drude-like dependence $\omega^{-2}$. At frequencies $\hbar \omega \ll k_B T$, the pre-factor before the collision integral in Eq.~(\ref{Ktot}) yields the $\omega^{-2}$-dependence. At large frequencies $\hbar \omega \geq k_B T$, the pre-factor behaves as $\omega^{-3}$, but the "collision integrals" $I_{ee,\omega}$ and $I_{eh,\omega}$ grow linearly due to the increasing number of final states.

\begin{figure}[ht]
\center{\includegraphics[width=0.95\linewidth]{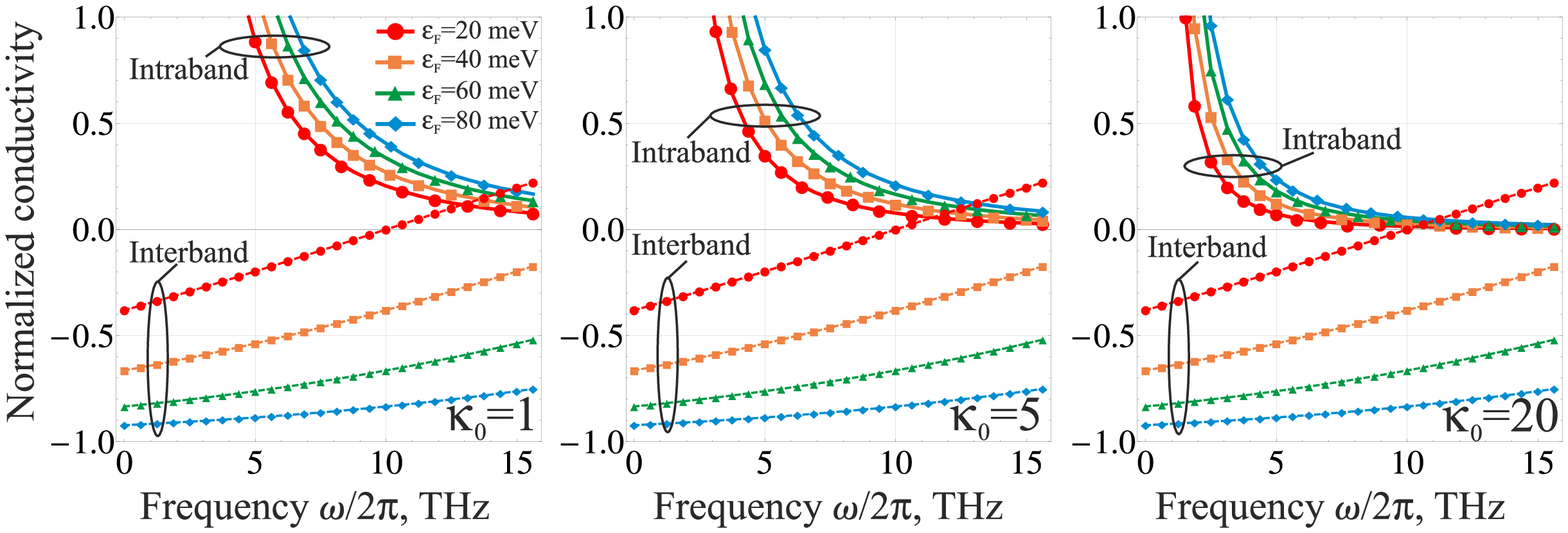}}
\caption{Real parts of the intraband (upper panels) and interband (lower panels) contributions, ${\rm Re}\sigma_{intra}$
and ${\rm Re}\sigma_{inter}$, to
dynamic conductivity 
normalized by $\sigma_q$ at different quasi-Fermi energies $\epsilon_F$ in graphene structures with different background dielectric constants $\kappa_0$ ($T = 300$~K).
}
\label{F2}
\end{figure}

In what follows, we shall study the dynamic conductivity of sufficiently clean pumped graphene. In this situation, the positive intraband conductivity is due to carrier-carrier collisions only, thus $\mathrm{Re} \sigma_{intra} \equiv \mathrm{Re} \sigma_{cc}$. Figure~\ref{F2} shows the dependences of $\mathrm{Re} \sigma_{intra}$ given by Eq.~(\ref{Ktot}) (normalized by $\sigma_q$) on frequency $\omega/2\pi$ for the symmetrically pumped graphene at different quasi-Fermi energies $\epsilon_F$ and different background dielectric constants $\kappa_0$ at room temperature $T=300$ K. In Fig.~ \ref{F2}, we also show the frequency-dependent real part of the dynamic conductivity associated with the interband transitions~\cite{Negative-dynamic-conductivity}: 

\begin{equation}
\label{Kinter}
\mathrm{Re} \sigma_{inter} = \sigma_q \tanh\left( \frac{\hbar \omega/2 - \epsilon_F}{2 k_B T}\right).
\end{equation}

From Fig.~\ref{F2} one can see that at frequencies above $\sim6.5$ THz at $\kappa_0=1$ (and above $\sim2.5$ THz at $\kappa_0 = 20$) the Drude conductivity due to c-c collisions lies below $\sigma_q $. Accordingly, the net conductivity $\mathrm{Re}(\sigma_{intra} + \sigma_{inter})$ can be negative in this frequency range at some level of pumping.

As the background dielectric constant $\kappa_0$ is increased, the intraband conductivity due to c-c collisions drops, which is illustrated in Fig.~\ref{F2}. {\it Ex facte}, one could expect that it scales as $\kappa_0^{-2}$, and increasing the dielectric constant one could reduce the radiation absorption due to c-c collisions almost to zero. Such considerations are actually irrelevant due to screening. The Thomas-Fermi screening momentum in pumped graphene is given by

\begin{equation}
q_{TF}\approx 8 \alpha_c \frac{k_B T}{v_F} \ln\biggl[1+ \exp\biggl({\frac{\epsilon_F}{k_B T}}\biggr)\biggr].
\end{equation}
Thus, at small momenta $q$ the Coulomb scattering matrix element is independent of $\kappa_0$. At large momenta $q \approx \epsilon_F/v_F$ the scattering matrix element $V^2_C(k_F) \propto [\kappa_0^2 + 8 e^2/(\hbar v_F)]^{-2}$, which slightly depends on $\kappa_0$ due to large value of bare coupling constant $e^2/\hbar v_F \approx 2.2$.

\begin{figure}[t]
\center{\includegraphics[width=0.95\linewidth]{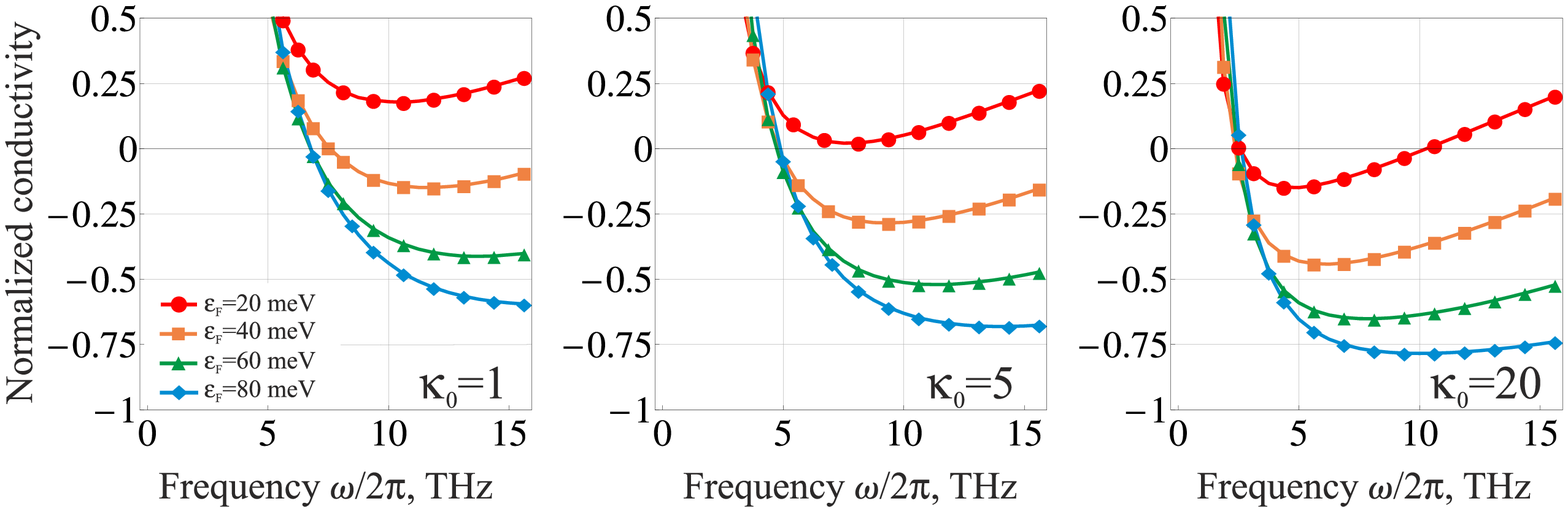}}
\caption{Real parts of net dynamic conductivity ${\rm Re}(\sigma_{intra} + \sigma_{inter})$ 
normalized by $\sigma_q$ at different quasi-Fermi energies $\epsilon_F$ in graphene structures with different background dielectric constants $\kappa_0$ ($T = 300$~K). }
\label{F3}
\end{figure}
In Fig.~\ref{F3} we show the net dynamic conductivity including interband and intraband c-c contributions. At weak pumping ($\epsilon_F \simeq 20$ meV) the negative dynamic conductivity is attainable only in graphene clad between high-$\kappa$ substrates. At elevated pumping ($\epsilon_F \simeq 40$ meV) the negative conductivity is possible at any value of $\kappa_0$. Further increase in the pumping level is not much efficient as the interband conductivity reaches its maximum value of $\sigma_q $ at $\epsilon_F \gg \hbar \omega$, while the intraband conductivity continues to grow with increasing $\epsilon_F$.

\begin{figure}[t]
\center{\includegraphics[width=0.95\linewidth]{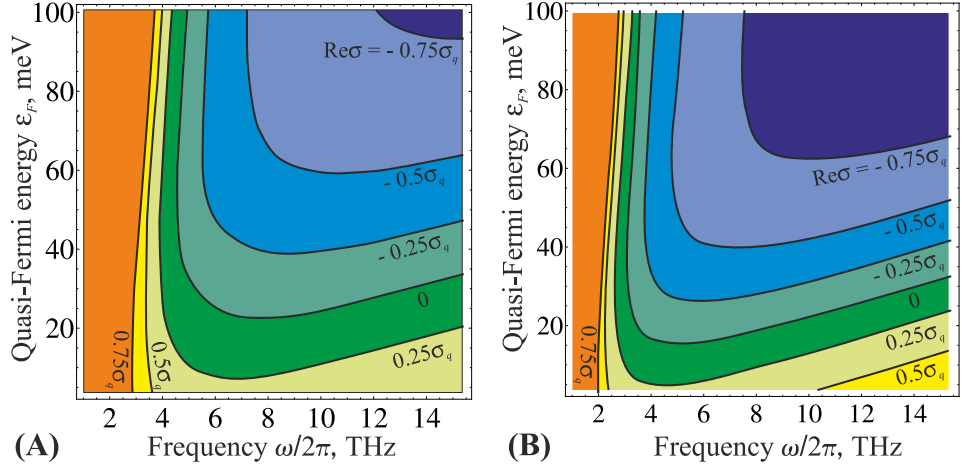}}
\caption{Color map of real part of net dynamic conductivity $\mathrm{Re} (\sigma_{intra} + \sigma_{inter})/\sigma_q$ vs frequency and quasi-Fermi energy for $\kappa_0=5$: (A) at $T=300$~K and (B) at $T=200$~K. The area $\mathrm{Re} (\sigma_{intra} + \sigma_{inter})/\sigma_q < 0.75$ is filled in solid color}
\label{F4}
\end{figure}
Both the interband and intraband c-c conductivities are actually functions of two dimensionless combinations $\hbar\omega/k_B T$ and $\epsilon_F/k_B T$. Once the temperature is reduced, $\mathrm{Re}\sigma_{intra}$ drops as well due to reduced phase space for collisions near Fermi-surface. The dependence $\mathrm{Re}\sigma_{inter}(\omega)$ becomes more abrupt in the vicinity of interband threshold $\hbar \omega = 2\epsilon_F$. Thereby, cooling of graphene sample is advantageous for achieving the negative dynamic conductivity. This is confirmed
by Fig.~\ref{F4}. In Figs.~4 (A) and 4 (B), we show the real part of the net dynamic conductivity versus frequency $\omega/2\pi$ and quasi-Fermi energy $\epsilon_F$ (values of $\mathrm{Re}\sigma$ are marked by colors) at the room temperature $T = 300$ K (A) and at $T=200$ K (B). It is worth noting that the temperature $T$ in the above equations is in fact
the effective temperature of the electron-hole system.
Therefore, 
the reduction of this temperature can be achieved not only by 
decreasing of the ambient temperature, but also by direct cooling of this system
at a certain pumping conditions.
In particular, the cooling of electron-hole system in the optically and electrically pumped graphene lasing structures due to the emission of optical phonons can be substantial ~\cite{Electrical-pumping-laser1,Opt.cooling}. 

\begin{figure}[t]
\center{\includegraphics[width=0.95\linewidth]{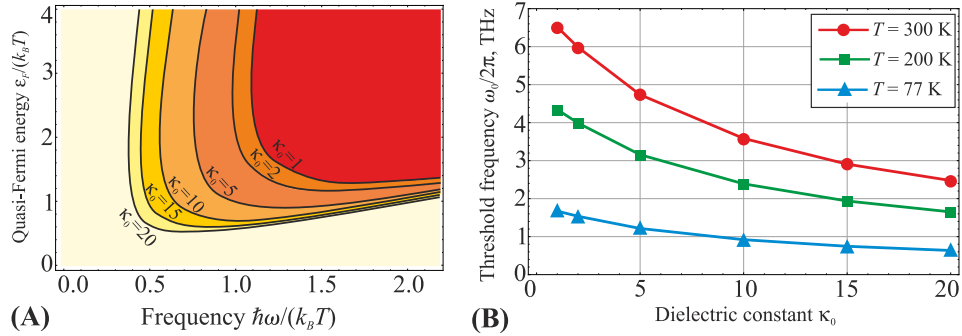}}
\caption{(A) Thresholds of negative dynamic conductivity at different values of background dielectric constant $\kappa_0$. Solid lines correspond to $\mathrm{Re}(\sigma_{intra} + \sigma_{inter})=0$. (B) Threshold frequencies $\omega_0/2\pi$ vs background dielectric constant at different temperatures $T$ ($\epsilon_F = 3 k_B T$)}
\label{F5}
\end{figure}
In Fig.~{\ref{F5}} (A) we show the threshold lines of the negative net dynamic conductivity [which are solutions of the equation $\mathrm{Re}\sigma(\omega,\epsilon_F)=0$] at different values of background dielectric constant. The normalized threshold frequency $\hbar \omega_0/k_B T$ slowly moves to lower values as $\kappa_0$ increases. This is illustrated in Fig.~\ref{F4} (B). At room temperature and $\kappa_0 = 1$, the negative conductivity could be attained only at frequencies $\omega/2\pi \geq 6.5$ THz. For graphene on SiO$_2$ substrate [$\kappa_0 = (\kappa_{\mathrm{SiO}_2}+1)/2 =2.5$], the room temperature threshold lies at $\omega_0/2\pi \approx 6$ THz. As discussed before, such slow decrease in threshold frequency is owing to the Thomas-Fermi screening of Coulomb interaction. Note that screening can be further enhanced by placing graphene close to the metal gate. In gated graphene, the Fourier component of Coulomb scattering potential $V_C(q)$ gains an additional factor $[1-\exp(-2qd)]$, where $d$ is the distance to the gate.

\begin{figure}[t]
\center{\includegraphics[width=0.60\linewidth]{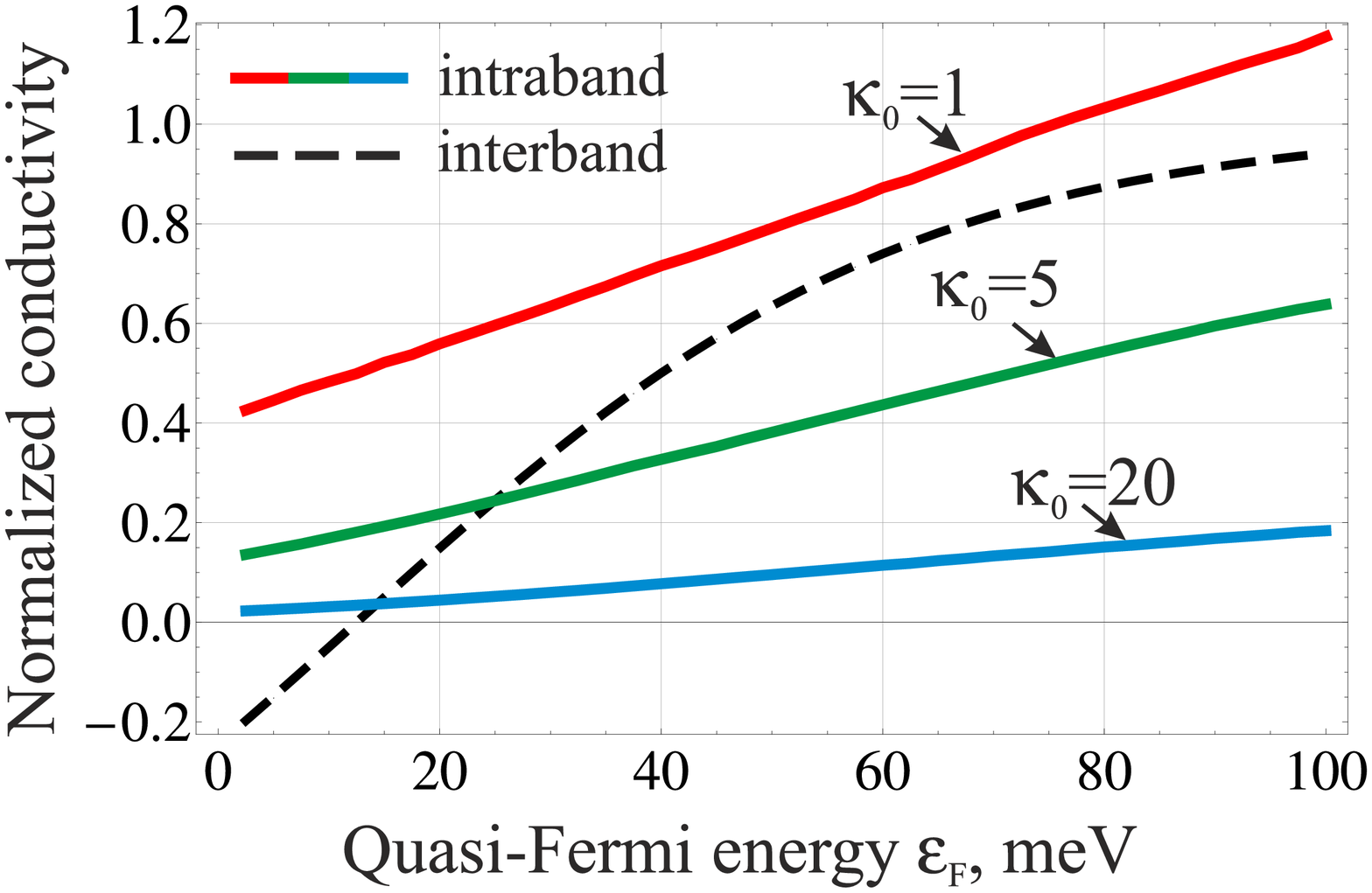}}
\caption{Real parts of intraband dynamic conductivity ${\rm Re}\sigma_{intra}$ (solid lines)
and interband conductivity $-{\rm Re}\sigma_{inter}$ 
(dashed line) as functions of quasi-Fermi energy $\epsilon_F$ at fixed frequency $\omega/(2\pi)=6$ THz and different values of background dielectric constant $\kappa_0$.}
\label{F6}
\end{figure}
The dependences of the inter- and intraband (carrier-carrier) conductivities on the quasi-Fermi energy at fixed frequency are shown in Fig.~\ref{F6}. The interband conductivity reaches its ultimate value of $\sigma_q$ at $\epsilon_F \gg \hbar \omega/2$. Intuitively, $\mathrm{Re}\sigma_{intra}$ and the corresponding radiation absorption coefficient should be proportional to the number of particle-particle collisions in unit time. This number is proportional to the carrier density squared, at least for classical carrier systems. In reality, the dependence of $\mathrm{Re}\sigma_{intra}$ on the quasi-Fermi energy $\epsilon_F$ appears to be almost linear. This, first of all, results from the screening and also from the restricted phase space for collisions in Fermi-systems in a narrow layer near the Fermi-surface.

\section{Discussion of the results}

The obtained relation for the intraband conductivity associated with the c-c scattering [Eq. (\ref{Ktot})] corresponds to a Drude-like dependence $\mathrm{Re}\sigma_{intra}\propto \omega^{-2}$. The latter tends to infinity at small frequencies. This tendency should be cut off at $\omega \sim \nu$, where $\nu$ is the electron collision frequency. At such small frequencies, the finite lifetime of quasiparticles leads to inapplicability of the Fermi golden rule. However, one can estimate the carrier-carrier collision frequency and the corresponding low-frequency conductivity by solving the kinetic equation for the massless electrons in graphene. We have generalized the variational approach to the solution of kinetic equation~\cite{Ziman} developed in Refs.~\cite{Kashuba,Quantum-critical} for intrinsic graphene, to the case of symmetrically pumped graphene. As a result, the following low-frequency dependence of intraband conductivity due to c-c scattering was obtained: 

\begin{equation}
\label{Sigma_cc}
\mathrm{Re}\sigma_{intra} = \frac{2 e^2}{\pi \hbar} \frac{k_B T}{\hbar} \ln \left[ 1 + \exp\left(\frac{\epsilon_F }{k_B T}\right) \right] \frac{ \nu_{cc}}{\omega^2 + \nu_{cc}^2},
\end{equation}
where the carrier-carrier collision frequency $\nu_{cc}$ includes the contributions from the
e-e and e-h scattering

\begin{equation}
\label{Nu_cc}
\nu_{cc} = \frac{\alpha_c^2}{4\pi^2} \frac{k_B T}{\hbar} \frac{ I_{ee,0} + I_{eh,0} }{\ln\biggl[1 + \exp\displaystyle\biggl(\frac{\epsilon_F }{k_B T}\biggr) \biggr]},
\end{equation}
and the ''zero'' index at dimensionless collision integrals means that they are taken at $\omega = 0$.

At intermediate frequencies $\nu_{cc} \ll \omega \leq k_B T/\hbar$, the results obtained from the kinetic equation[Eq. (\ref{Sigma_cc})] and the Fermi golden rule [Eq.~(\ref{Ktot})] coincide. At room temperature, zero Fermi energy, and background dielectric constant $\kappa_0 = 1$, the characteristic collision frequency is equal to $\nu_{cc}\approx 10^{13}$ s$^{-1}$. Once $\kappa_0$ or $\epsilon_F$ are increased, the collision frequency decreases due to screening of the Coulomb potential. Hence, for the 'target' frequencies $\omega/(2\pi) = 5 - 10$ THz one can readily use the golden rule result [Eq.~(\ref{Ktot})] to evaluate the carrier-carrier scattering contribution to the optical conductivity.

Until now we considered the intraband conductivity due to the c-c collisions only, as if other scattering processes were removed. In realty, the acoustic phonon contribution is unavoidable as well. The corresponding collision frequency at $\epsilon_F \gg k_B T$ can be presented as $\nu_{ph} = \nu_0 (\epsilon_F/k_BT)$, where $\nu_0 \simeq 6\times 10^{11}$ s$^{-1}$~\cite{Vasko-Ryzhii}. Thus, at least for not very large values of $\kappa_0$, the c-c scattering contribution to intraband conductivity appears to be larger than the acoustic phonon contribution.

In our estimates for the Coulomb scattering probability, we have used the static (Thomas-Fermi) approximation for dielectric function of graphene. The use of the dynamic dielectric function can be mandatory to avoid the divergence in collision integrals for collinear scattering~\cite{Collinear-Scattering}. However, in Eqs.~(\ref{Iee}) and (\ref{Ieh}) such a divergence does not appear due to the presence of factor $\Delta {\bf n}^2$. The difference between $\mathrm{Re}\sigma_{intra}$ obtained using static and dynamic screening is below 10 \%. 

\begin{figure}[t]
\center{\includegraphics[width=0.95\linewidth]{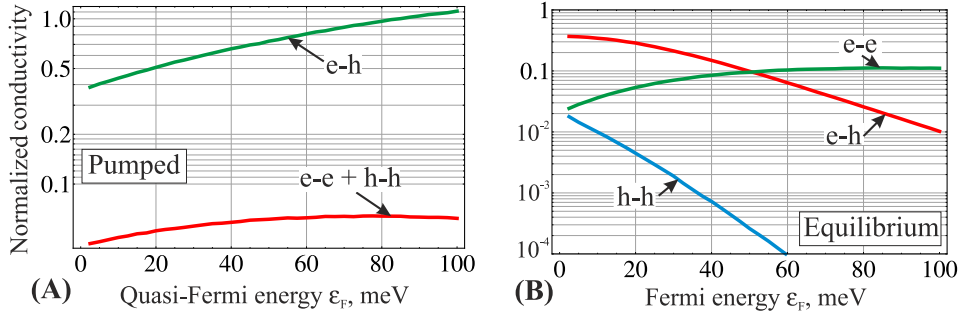}}
\caption{Separate contributions of e-h, e-e, and h-h scattering processes to real part of interband conductivity: (A) for pumped graphene and (B) for electron-doped graphene in thermodynamic equilibrium ($\kappa_0 = 1 $, $\omega/2\pi = 6$~THz, and $T = 300$~K).}
\label{F7}
\end{figure}

As known, the e-e and h-h scattering does not lead to the finite resistivity and the absorption of radiation by the carriers with the parabolic dispersion law (in the absence of umklapp processes~\cite{Ziman}). In graphene, such an effect is possible due to the momentum-velocity decoupling. Formally, e-e collisions in pumped graphene are of the same ''strength'' as e-h collisions [see Eqs.~(\ref{Kee}) and (\ref{Keh})]. However, the numerical comparison of Eqs.~(\ref{Kee}) and (\ref{Keh}) shows that the e-e and h-h collisions play a minor role in the radiation absorption in the pumped graphene. The major contribution to $\mathrm{Re}\sigma_{cc}$ comes from the electron-hole collisions, while $\mathrm{Re}\sigma_{ee}$ actually constitutes less than 10\% of $\mathrm{Re}\sigma_{cc}$ (see Fig.~\ref{F7} A). This can be understood by analyzing the change in current in a single collision act. The corresponding change is given by the terms $(\Delta {\bf n}_{ee})^2$ and $(\Delta {\bf n}_{eh})^2$ in Eqs.~(\ref{Iee}) and (\ref{Ieh}). Expanding in powers of ${\bf q}$ and averaging over directions, one obtains $(\Delta {\bf n}_{ee})^2 \propto q^2(k_1^{-1} - k_2^{-1})^2$ and $(\Delta {\bf n}_{eh})^2 \propto q^2(k_1^{-1} + k_2^{-1})^2$. The latter implies that collisions of electrons with sufficiently different energies only play significant role in the e-e and h-h scattering-assisted absorption. For electron-hole collisions there is no such a restriction.

For the electron-hole system in equilibrium, the relative contributions of e-e and e-h scattering to the intraband conductivity are much different [see Fig.~\ref{F7}(B)]. The h-h and e-h contributions drop quickly with increasing the Fermi energy due to vanishingly small number of holes. The remaining part of intraband conductivity due to the e-e collisions is of order of $0.1\sigma_q$ at $\omega/(2\pi) = 6$ THz and $\kappa_0=1$. The corresponding contribution to radiation absorption can be in principle detected for rather clean samples.


Recent experiments have shown that the THz radiation gain in the pumped graphene can greatly exceed $2.3$ \%~\cite{SPP-enhancement-experim}. This phenomenon is attributed to the self-excitation of surface plasmon-polaritons in electron-hole system with population inversion~\cite{SPP-enhancement-theory}. The threshold of plasmon self-excitation in the pumped graphene is also governed by the ratio of the negative interband and positive intraband conductivities. The latter results mainly from the e-h collisions~\cite{Our-hydrodynamics}. The presented calculations show that terahertz plasmons are rather amplified than damped in the pumped graphene, at least at high pumping levels and large background dielectric constants.

\section{Conclusions}

We have shown that in graphene there exists a strong mechanism of the intraband Drude-like radiation absorption, assisted by the carrier-carrier collisions. In such a process, a carrier absorbs a photon and transfers the excess energy and momentum to the other carrier, electron or hole. The radiation absorption assisted by the electron-electron and hole-hole collisions is possible only in semiconductors with non-parabolic bands, and graphene (with its linear energy spectrum of electrons and holes) is one of brightest representatives of such a family. We have evaluated the carrier-carrier scattering contribution to the intraband conductivity using the second-order perturbation theory and the Fermi golden rule. As demonstrated, the radiation absorption due to the carrier-carrier collisions can markedly surpass the Drude absorption due to the scattering on thermal or static disorder. This is especially pronounced in the pumped graphene with a large number of carriers in both bands. 

We have studied the effect of carrier-carrier scattering on the conditions of the negative dynamic conductivity in pumped graphene by comparing the positive intraband dynamic conductivity due to the carrier-carrier collisions and the negative interband conductivity. We have shown that at room temperature, the net negative dynamic conductivity is attainable at frequencies above $\sim6.5$ THz in suspended graphene. This frequency threshold can be shifted to lower values ($\sim2.5$ THz at room temperature) for graphene on high-$\kappa$ substrates, as the presence of dielectric partially screens the Coulomb interaction. Cooling of the electron-hole system also reduces the negative conductivity threshold. Our results show that screening of Coulomb interaction by carriers in graphene itself (Thomas-Fermi screening) significantly affects the real part of intraband conductivity. Due to the screening, the c-c contribution to the intraband conductivity grows slowly (almost linearly) with increasing quasi-Fermi energy of nonequilibrium electrons and holes.

\section*{Acknowledgments}
The authors are grateful to M.Ryzhii for useful comments. The work at RIEC was supported by the Japan Society for Promotion of Science (Grant-in-Aid for Specially Promoting Research $\#$ 23000008), Japan. The work at IPT RAS was supported by the Russian Foundation of Basic Research (grant $\#$ 14-07-00937).
The work by D.S was also supported by the grant of the Russian Foundation of Basic Research $\#$ 14-07-31315 and by the JSPS Postdoctoral Fellowship For Foreign Researchers (Short-term), Japan.

\setcounter{equation}{0}
\renewcommand{\theequation} {A\arabic{equation}}
\section*{Appendix. Evaluation of Coulomb integrals}

We introduce the dimensionless energies of incident particles $\varepsilon_1 = k_{1+}$ and $\varepsilon_2 = k_{2-}$, and the transferred energy $\delta \varepsilon = k_{1+} - k_{1-} - w/2 = k_{2+} - k_{2-} - w/2$. Here, for brevity, $w = \hbar \omega /(k_B T)$. The kernel of
the e-e Coulomb integral depends only on those energies and the transferred momentum~$Q$:

$$
C_{ee}\left( {\varepsilon_1},{\varepsilon_2},\delta \varepsilon ,Q \right) = \int d{\bf k} _1 d{\bf k}_2 
\left( \Delta {\bf n}_{ee} \right)^2 
\cos^2 (\theta_{1\pm}/2) \cos^2(\theta_{2\pm}/2) \\
\delta \left( \varepsilon_1 - k_{1+} \right) \delta \left( \varepsilon_2 - k_{2-} \right) 
$$
\begin{equation}
\label{Cee}
\times\delta \left[ \delta \varepsilon - \left( k_{2+} - k_{2-} - \omega/2 \right) \right] 
\delta \left[ \delta \varepsilon - ( k_{1+} - k_{1-} +\omega/2 ) \right].
\end{equation}
In these notations, $I_{ee,\omega}$ is rewritten as
$$
I_{ee,\omega}=
\int\limits_{w/2}^{\infty }\frac{QdQ}{\left( Q+Q_{TF} \right)^2} 
\int\limits_{\delta\varepsilon_{\min}}^{\delta\varepsilon_{\max}} d\delta \varepsilon 
\int\limits_{\varepsilon_{1\min}}^{\infty } d \varepsilon_1 
\int\limits_{\varepsilon_{2\min}}^{\infty } d \varepsilon_2 
C_{ee}\left( {\varepsilon_1},{\varepsilon_2},\delta \varepsilon ,Q \right) \\
F( \varepsilon_1 ) F( \varepsilon_2 ) 
$$
\begin{equation}
\label{Iee-App}
\times
\left[ 1 - F( \varepsilon_2 + \delta \varepsilon + w/2) \right] 
\left[ 1 - F( \varepsilon_1 - \delta \varepsilon + w/2) \right].
\end{equation}
The limits of integration will be found further. In elliptic coordinates the Coulomb kernel is evaluated analytically. The change of variables is ($i = 1,2$)
\begin{equation}
{\bf k}_{i}=\frac{Q}{2}\left\{ \cosh u_i\cos v_i, \sinh u_i \sin v_i \right\},
\end{equation}
where $u_i > 0$, and $-\pi < v_i < \pi$. The product of four delta-functions in Eq.~(\ref{Cee}) can be rewritten as
$$
\frac{4}{Q^4}
\delta \biggl( \cos v_2-\frac{\delta \varepsilon + w/2}{Q} \biggr)
\delta \biggl( \cos v_1-\frac{\delta \varepsilon - w/2}{Q}\biggr) 
\delta \biggl( \cosh u_1-\frac{w/2 - \delta \varepsilon +2\varepsilon_1}{Q}\biggr)
$$
\begin{equation}\label{App-deltas}
\times\delta \biggl( \cosh u_2-\frac{w/2 + \delta \varepsilon +2\varepsilon_2}{Q} \biggr).
\end{equation}
Considering the arguments of delta-functions in Eq.~(\ref{App-deltas}), one finds the limits of integration over energies in Eq.~(\ref{Iee-App}) from general requirements $\cosh u_i > 1$, $|\cos v_i|<1$, $i=1,2$:
\begin{equation}
{\varepsilon_{1}} > \frac{( Q + \delta \varepsilon - w/2 )}{2};
\, 
{\varepsilon_{2}} > \frac{( Q - \delta \varepsilon - w/2)}{2},
\,
Q > w/2;
\, -Q + w/2 < \delta \varepsilon < Q - w/2.
\end{equation}
Integration over $du_1$, $du_2$, $dv_1$, and $dv_2$ removes the delta-functions, and the Coulomb kernel becomes

\begin{equation}
\label{Coulomb-kernel}
C( {\varepsilon_{1}},{\varepsilon_{2}},\delta \varepsilon ,Q)
=
\frac{4(\Delta \bf{n})^2}{Q^4}
\frac{\cos^2 (\theta_{1\pm}/2) \cos^2(\theta_{2\pm}/2) }
{\left| \sin v_1 \sin v_2 \sinh u_1 \sinh u_2 \right|}
k_{1+}k_{1-}k_{2+}k_{2-}.
\end{equation}
Here the $\sin$- and $\sinh$-terms in the denominator appear from the rule $\delta \left[f(x)\right] = \sum_i |f'(x_i)|^{-1}\delta(x-x_i)$, where the summation is performed over all zeros $x_i$ of $f(x)$. In Eq.~(\ref{Coulomb-kernel}) $k_{1+}k_{1-}k_{2+}k_{2-}$ is the Jacobian determinant for elliptic coordinates.

The envelope function overlap factors are also concisely expressed in the elliptic coordinates:

$$
\cos^2 (\theta_{1\pm}/2) \cos^2(\theta_{2\pm}/2) k_{1+}k_{1-}k_{2+}k_{2-} 
$$
\begin{equation}
= 
\frac{1}{16}
\left[ {\left( 2\varepsilon_1 - \delta \varepsilon + w/2 \right)^2} - Q^2 \right]
\left[ {\left( 2\varepsilon_2 + \delta \varepsilon + w/2 \right)^2} - Q^2 \right].
\end{equation}

As a result, the Coulomb kernel takes on the following form:

$$
{{C}_{ee}}\left( {\varepsilon_{1}},{\varepsilon_{2}},\delta \varepsilon ,Q \right)=\frac{\left( \Delta {\bf n} \right)_{ee}^2}{4}\frac{\sinh u_1 \sinh u_2}{\sin v_1 \sin v_2}
$$
\begin{equation}
\label{Final-Kernel}
=\frac{( \Delta {\bf n} )_{ee}^2}{4}\sqrt{\frac{\left[ \left( 2 \varepsilon_1 - \delta \varepsilon +w/2 \right)^2 - Q^2 \right]\left[ ( 2\varepsilon_2 + \delta \varepsilon + w/2)^2 - Q^2 \right]}{\left[ Q^2 - (\delta \varepsilon + w/2)^2 \right]\left[ Q^2 - ( \delta \varepsilon -w/2 )^2 \right]}}.
\end{equation}
The expression for $( \Delta {\bf n} )_{ee}^2$ in elliptic coordinates is quite lengthy, here we present in only in the limit $w\to 0$:
$$
( \Delta {\bf n} )_{ee}^2 = \frac{2 \left(Q^2-\delta \varepsilon^2\right)} {\varepsilon_1 \varepsilon_2 ( \varepsilon_1-\delta \varepsilon) (\delta \varepsilon +\varepsilon_2 )}
$$
\begin{equation}
\times \left[\delta \varepsilon^2+4 \delta \varepsilon (\varepsilon_2-\varepsilon_1 ) + 2 (\varepsilon_1-\varepsilon _2)^2 - \frac{\delta \varepsilon^2 (\delta\varepsilon -2\varepsilon_1 ) ( \delta \varepsilon + 2 \varepsilon_2)}{Q^2} \right].
\end{equation}
Note that Eq.~(\ref{Final-Kernel}) contains the denominator divergent at $w \to 0$, which manifests the collinear scattering anomaly for massless particles~\cite{Quantum-critical,Kashuba,Collinear-Scattering}. This divergence is removed by the similar term in the expression for $( \Delta {\bf n} )_{ee}^2$, which stands in the numerator.


\begin{thebibliography}{99}

\bibitem{Graphene-photonics-review-1}
F. Bonaccorso, Z. Sun, T. Hasan, A.C. Ferrari, ''Graphene photonics and optoelectronics,'' Nature Photonics {\bf 4}, 611 (2010).

\bibitem{Graphene-photonics-review-2}
A. Tredicucci and M. S. Vitiello, ''Device concepts for graphene-based terahertz photonics,'' IEEE J. Sel. Top.
Quantum Electron. 20, 8500109 (2014).

\bibitem{Pumped-graphene-materials}
P. Weis, J. L. Garcia-Pomar, M. Rahm ''Towards loss compensated and lasing terahertz metamaterials based on optically pumped graphene,'' Opt. Express {\bf 22} Iss. 7, 8473 (2014).

\bibitem{Optical-pumping-laser}
A. Dubinov, V.Ya. Aleshkin, M. Ryzhii, T. Otsuji, V. Ryzhii, ''Terahertz laser with optically pumped graphene layers and Fabri-Perot resonator,'' Appl. Phys. Express {\bf 2}, 092301 (2009).

\bibitem{Electrical-pumping-laser1}
V. Ryzhii, M. Ryzhii, V. Mitin, T. Otsuji, ''Toward the creation of terahertz graphene injection laser,'' J. Appl. Phys. {\bf 110}, 094503 (2011).

\bibitem{Electrical-pumping-laser2}
V. Ryzhii, A. Dubinov, T. Otsuji, V.Ya. Aleshkin, M. Ryzhii, and M. Shur, ''Double-graphene-layer terahertz laser: concept, characteristics, and comparison,'' Opt. Express {\bf 21} Iss. 25, 31567 (2013).

\bibitem{THz-amplification-experim}
S. Boubanga-Tombet, S. Chan, T. Watanabe, A. Satou, V. Ryzhii, T. Otsuji, ''Ultrafast carrier dynamics and terahertz emission in optically pumped graphene at room temperature,'' Phys. Rev. B {\bf 85}, 035443 (2012).

\bibitem{THz-gain-loss-QCL}
M. Martl, J. Darmo, C. Deutsch, M. Brandstetter, A. M. Andrews, P. Klang, G. Strasser, and K. Unterrainer, ''Gain and losses in THz quantum cascade laser with metal-metal waveguide,'' Opt. Express {\bf 19}, 733 (2011).

\bibitem{Collinear-Scattering}
D. Brida, A. Tomadin, C. Manzoni, Y. J. Kim, A. Lombardo, S. Milana, R. R. Nair, K. S. Novoselov, A. C. Ferrari, G.Cerullo, M. Polini ''Ultrafast collinear scattering and carrier multiplication in graphene,'' Nature Communications {\bf 4}, Article number: 1987 (2013).

\bibitem{Optical-conductivity-measurement}
K.F. Mak, M.Y. Sfeir, Y. Wu, C.H. Lui, J.A. Misewich, and T.F. Heinz, ''Measurement of the optical conductivity of graphene,'' Phys. Rev. Lett. {\bf 101}, 196405 (2008).

\bibitem{Negative-dynamic-conductivity}
V. Ryzhii, M. Ryzhii and T. Otsuji, ''Negative dynamic conductivity of graphene with optical pumping,'' J. Appl. Phys. {\bf 101}, 083114 (2007).

\bibitem{Threhold-of-population-inversion}
A. Satou, V. Ryzhii, Y. Kurita, T. Otsuji ''Threshold of terahertz population inversion and negative dynamic conductivity in graphene under pulse photoexcitation,'' J. Appl. Phys. {\bf 113}, 143108 (2013).

\bibitem{Relaxation-kinetics}
J.M. Dawlaty, S. Shivaraman, M. Chandrashekhar, F.~Rana, and M.G. Spencer, ''Measurement of ultrafast carrier dynamics in epitaxial graphene,'' Appl. Phys. Lett. {\bf 92}, 042116 (2008).

\bibitem{Kashuba}
A.B. Kashuba, ''Conductivity of defectless graphene,'' Phys. Rev. B {\bf 78}, 085415 (2008).

\bibitem{Quantum-critical}
L. Fritz, J. Schmalian, M. M{\"u}ller, and S. Sachdev, ''Quantum critical transport in clean graphene,'' Phys. Rev. B {\bf 78}, 085416 (2008).

\bibitem{Our-hydrodynamics}
D. Svintsov, V. Vyurkov, S. Yurchenko, V. Ryzhii, T. Otsuji, ''Hydrodynamic model for electron-hole plasma in graphene,'' J. Appl. Phys. {\bf 111} (8), 083715 (2012).

\bibitem{Interactions-transport}
M. Sch{\"u}tt, P.M. Ostrovsky, I.V. Gornyi, and A.D. Mirlin, ''Coulomb interaction in graphene: Relaxation rates and transport,'' Phys. Rev. B. {\bf 83}, 155441 (2011).

\bibitem{Electron-interactions}
V. N. Kotov, B. Uchoa, V. M. Pereira, F. Guinea, and A.H.~Castro Neto, ''Electron-Electron Interactions in Graphene: Current Status and Perspectives,'' Rev. Mod. Phys. {\bf 84}, 1067 (2012).

\bibitem{Interplay-of-inter-and-intra}
F. T. Vasko, V. V. Mitin, V. Ryzhii, and T. Otsuji, ''Interplay of intra- and interband absorption in a disordered graphene,'' Phys. Rev. B {\bf 86}, 235424 (2012).

\bibitem{EE-absorption-in-QW}
G. G. Zegrya, V. E. Perlin, ''Intraband absorption of light in quantum wells induced by electron-electron collisions,'' Semiconductors {\bf 32}, 417 (1998).

\bibitem{Hwang-Das-Sarma}
E. H. Hwang and S. Das Sarma ''Dielectric function, screening, and plasmons in two-dimensional graphene,'' Phys. Rev. B {\bf 75}, 205418 (2007).

\bibitem{LL-Quantum-mechanics}
L. D. Landau and E.M. Lifshitz \textit{Quantum Mechanics} (Pergamon Press, 1965).

\bibitem{Opt.cooling}
V. Ryzhii, M. Ryzhii, V. Mitin, A. Satou, and T. Otsuji, "Effect of heating and cooling of photogenerated electron-hole plasma in optically pumped graphene on population inversion", Jpn. J. Appl. Phys. {\bf 50}, 094001 (2011).

\bibitem{Ziman}
J. M. Ziman, \textit{Electrons and Phonons} (Oxford University Press, 1960).

\bibitem{Vasko-Ryzhii}
F. T. Vasko and V. Ryzhii, ''Voltage and temperature dependencies of conductivity in gated graphene,'' Phys. Rev. B {\bf 76}, 233404 (2007).

\bibitem{SPP-enhancement-experim}
T. Watanabe, T. Fukushima, Y. Yabe, S.A. Boubanga Tombet, A. Satou, A. Dubinov, V.Ya. Aleshkin, V. Mitin, V. Ryzhii and T. Otsuji, ''The gain enhancement effect of surface plasmon polaritons on terahertz stimulated emission in optically pumped monolayer graphene,'' New J. Phys. {\bf 15}, 075003 (2013).

\bibitem{SPP-enhancement-theory}
A. Dubinov, V. Aleshkin, V. Mitin, T. Otsuji and V. Ryzhii, ''Terahertz surface plasmons in optically pumped graphene structures,'' J. Phys.: Condens. Matter {\bf 23}, 145302 (2011).

\end{thebibliography}
\end{document}